\documentclass[preprint]{aastex}
\usepackage{natbib}
\usepackage{subfigure}
\title{Simulation of the Formation of a Solar Active Region}

\author{M. C. M. Cheung\affil{Lockheed Martin Solar and Astrophysics Laboratory, Palo Alto, CA 94304, USA.}}
\author{M. Rempel\affil{High Altitude Observatory, NCAR, P.O. Box 3000, Boulder, Colorado 80307, USA}}
\author{A. M. Title\affil{Lockheed Martin Solar and Astrophysics Laboratory, Palo Alto, CA 94304, USA.}}
\author{M. Sch\"ussler\affil{Max Planck Institute for Solar System Research, Katlenburg-Lindau, 37191, Germany.}}

\begin{abstract}
We present a radiative magnetohydrodynamics simulation of the formation of an Active Region on the solar surface. The simulation models the rise of a buoyant magnetic flux bundle from a depth of 7.5 Mm in the convection zone up into the solar photosphere. The rise of the magnetic plasma in the convection zone is accompanied by predominantly horizontal expansion. Such an expansion leads to a scaling relation between the plasma density and the magnetic field strength such that $B\propto\varrho^{1/2}$. The emergence of magnetic flux into the photosphere appears as a complex magnetic pattern, which results from the interaction of the rising magnetic field with the turbulent convective flows. Small-scale magnetic elements at the surface first appear, followed by their gradual coalescence into larger magnetic concentrations, which eventually results in the formation of a pair of opposite polarity spots. Although the mean flow pattern in the vicinity of the developing spots is directed radially outward, correlations between the magnetic field and velocity field fluctuations allow the spots to accumulate flux.~{Such correlations result from the Lorentz-force driven, counter-streaming motion of opposite-polarity fragments}. The formation of the simulated Active Region is accompanied by transient light bridges between umbrae and umbral dots. Together with recent sunspot modeling, this work highlights the common magnetoconvective origin of umbral dots, light bridges and penumbral filaments.
\end{abstract}

\begin{document}

\section{Introduction}
Active regions  (ARs) are the most prominent manifestation of the large scale solar
magnetic field in the photosphere of the Sun. Understanding the underlying
flux emergence process is a crucial step toward understanding the connection
between the dynamo processes in the solar convection zone and magnetic activity
in the photosphere and above.

Observations of active regions on the solar surface indicate that, in terms of magnetic flux content, there is a continuous spectrum of active region sizes~\citep{Zwaan:AppearanceOfMagneticFlux,Zwaan:ElementsAndPatterns,Harvey:EphemeralActiveRegions,SchrijverZwaan:MagneticActivity,Hagenaar:Ephemeralregions,HagenaarSchrijver:BipolarRegions}. At the small end of the spectrum,~\emph{ephemeral ARs} (which have a flux of $3\times10^{18} - 1\times10^{20}$ Mx) manifest themselves in the solar photosphere in terms of mixed polarity field and  magnetic bright points but not necessarily pores~\citep{Cheung:SolarSurfaceEmergingFluxRegions}.~\emph{Small ARs}, which have fluxes of $1\times 10^{20}$ to $5\times 10^{21}$ Mx in each polarity, contain solar pores.~\emph{Large ARs}, which have even more magnetic flux (up to $4\times 10^{22}$ Mx), contain sunspots with penumbrae.

Since the original suggestion by~\citet{Parker:FormationOfSunspots}, there has been a large body of theoretical and observational work supporting the scenario that sunspots (and ARs) form as a result of the buoyant rise of magnetic flux bundles from the solar convection zone to the solar atmosphere. On the observational side, the work of~\citet{Zwaan:AppearanceOfMagneticFlux,Zwaan:ElementsAndPatterns} and~\citet{StrousZwaan:SmallScaleStructure} showed that ARs form as a consequence of a succession of small-scale flux emergence events on the solar surface. The observational studies in these papers noted that although magnetic flux emerges as small bundles, these bundles subsequently migrate and coalesce in a systematic fashion, eventually leading to formation of pores and sunspots (i.e. patches of predominantly one polarity). Recently, an analysis of spectropolarimetric observations by~\citet{Schlichenmaier:PenumbraFormation} shows how the accumulative of flux by a solar pore resulted in the formation of penumbral filaments.

Numerical magnetohydrodynamics (MHD) simulations of the rise of buoyant magnetic flux tubes have yielded a wealth of information regarding the flux emergence process. For example, numerical models of the rise of toroidal flux tubes from the bottom of the convection zone based on the thin flux tube approximation~\citep{RobertsWebb:VerticalMotions,Spruit:ThinFluxTube} are able to reproduce global properties of ARs such as their tilt angles~\citep{DSilva:ARTiltAngle}, their emergence latitudes~\citep{Caligari:EmergingTubesPartI,Caligari:EmergingTubesPartII} and asymmetries between the leading and following polarities~\citep{Fan:ThinFluxTubeI,Moreno-Insertis:pf-assymetries,Caligari:EmergingTubesPartI,Caligari:EmergingTubesPartII}. Due to the invalidity of the thin flux tube approximation in the uppermost ten or so Mm of the solar convection zone, such studies have limited applicability for examining how ARs form on the solar surface. Complementing this line of work, multidimensional MHD simulations~\citep[e.g.][]{Forbes:FluxEmergence,Isobe:Rayleigh-TaylorInstability,Shibata:FluxEmergence,Archontis:EmergenceIntoCorona,Manchester:EruptionOfEmergingFluxRope,Magara:EmergingFluxSurfaceFlows,Fang:FluxEmergence} into the solar atmosphere have greatly improved our understanding of how the buoyant rise of magnetic fields into the solar atmosphere affect local dynamics. For a historical perspective of such work, the reader is referred to the review articles by~\citet{Moreno-Insertis:Review},~\citet{Fisher:SolarDynamoAndEmergingFlux} and~\citet{Fan:LivingReview,Fan:LivingReview2009}. 

Recent models that include the treatment of convective motions and radiative heating/cooling (either by solving the radiative transfer equation or by parameterized terms in the energy equation) have shown these two effects to be very important to the properties of flux emergence~\citep{SteinNordlund:SmallScaleMagnetoconvection,Cheung:FluxEmergenceInGranularConvection,Cheung:SolarSurfaceEmergingFluxRegions,Isobe:ConvectionDrivenEmergence,Abbett:MagneticConnection,MartinezSykora:TwistedFluxEmergence,MartinezSykora:TwistedFluxEmergenceII,TortosaAndreu:FluxEmergence,Yelles:StokesDiagnostics,Fang:FluxEmergence}. Despite these advances,  a comprehensive
numerical model of the formation of an AR resulting from flux emergence has been lacking. This
is a consequence of the vast range of length and time scales encountered in the
convection zone as consequence of a density variation by about six orders of
magnitude from the base to the top. To cope with this difficulty numerical
models focus either on the deep convection zone leaving out the upper most
10 - 20 Mm or on the upper most 10 Mm including the photosphere.  

While the
former typically utilize the anelastic approximation~\citep{Abbett:3DTubeInConvection,Fan:3DTubeConvection,Jouve:3DFluxTube}~the latter have to be
based on fully compressible MHD. Regardless of the differences a common
challenge in both cases is to explain the formation of rather coherent
sunspots with several kG field strength from magnetic field which has risen
through a highly stratified atmosphere and consequently should have weakened considerably
through expansion. Addressing this aspect through a self-consistent
numerical simulation capturing the rise of magnetic flux from the
base of the convection zone into the photosphere is currently out of reach, however
much can be learned from a numerical simulation containing the last
$7.5$ Mm beneath the photosphere. While the geometric extent is only $3.5\%$ of
the convection zone depth, the density drop of $3$ orders of magnitude (corresponding to ten pressure scale-heights)
is comparable to the drop in the remaining $96.5\%$ of the convection zone.
In this paper we investigate the flux emergence process in the upper most
$7.5$ Mm  of the convection zone resulting from the advection of a semitorus-shaped flux tube across the bottom
boundary of our computational domain. 

\section{Simulation Setup}
The radiative MHD simulation was carried out with the~\emph{MURaM} code~\citep{Voegler:MURaM,Rempel:SunspotStructure}, which takes into account radiative energy transport in the energy equation (assuming local thermodynamic equilibrium) and a realistic equation of state. The MURaM code has been used to model a variety of magnetic structures in the solar photosphere and convection zone. The Cartesian domain spans $92$ Mm $\times$ $49$ Mm in the two horizontal directions and $8.2$ Mm in the vertical direction (with horizontal and vertical grid-spacing of $48$ and $32$ km respectively). The base of the photosphere ($z=0$, the mean geometrical height where the Rosseland optical depth $\tau_R$ has a value of unity) is located $7.5$ Mm above the bottom boundary of the simulation domain. Before the introduction of a magnetic flux tube, a purely hydrodynamic simulation was performed to allow the radiatively driven convection to relax to statistical equilibrium.

To mimic the rise of magnetic flux into the topmost layers of the convection zone, an axisymmetric, twisted flux tube with the shape of a semitorus is kinematically advected into the domain through the bottom boundary. Similar time-dependent boundary conditions to kinematically advect magnetic field into the top layers of the convection zone have previously been used for flux emergence simulations.~\citet{SteinNordlund:SmallScaleMagnetoconvection} and \citet{Stein:SolarFluxEmergenceSimulations} used a boundary condition which advect purely horizontal field (of uniform field strength and orientation) through upflow regions.~\citet{MartinezSykora:TwistedFluxEmergence} used a boundary condition which advected an horizontal twisted flux tube. The boundary condition for the current simulation is implemented by specifying the time-dependent fluxes of vertical mass, momentum, energy and magnetic field (i.e. terms acted upon by the divergence operator in the conservation form of the MHD equations) consistent with the rise of the torus. Following~\citet*{Fan:Arcade}, the magnetic field distribution of the flux tube in a spherical coordinate system with the torus centered at the origin is given by 
\begin{eqnarray}
\vec{\bf B} &=& \nabla \times [\frac{A(r,\theta)}{r\sin{\theta}}\hat{\phi}] + B_\phi (r,\theta)\hat{\phi},\\
A &=& \frac{1}{2}\lambda a B_t \exp{(-\omega^2/a^2)},{\rm~and} \\
B_\phi &=& (r \sin{\theta})^{-1}a B_t \exp{(-\omega^2/a^2)}. 
\end{eqnarray}
\noindent $\omega = (r^2+R^2 - 2rR\sin{\theta})^{1/2}$ is the distance of a point $\vec{\bf r}$ from the toroidal axis of the tube, and $a$ and $R$ are the semi-minor and -major radii of the torus, respectively. The dimensionless twist parameter $\lambda$ is equivalent to the parameter $a^{-1}q$ used by~\citet{Fan:Arcade}. For this simulation,  we took $\lambda=0.5$, $R = 16$ Mm and $a = 3.6$ Mm. $B_t = 94$ kG such that the toroidal field strength at the tube axis is $B_\phi(\omega=0) = a R^{-1}B_t = 21$ kG (corresponding to a plasma $\beta$ of $140$). For $\omega>\sqrt{2} a$, $\vec{\bf B}_{\rm tube}$ is set to zero. The total toroidal flux content of the tube is $\Phi_0 = 7.6\times10^{21}$ Mx. Averaged over the cross-section of the tube (i.e. $ 0 \le \omega \le \sqrt{2}a$), the mean toroidal field strength is $9$ kG. This is somewhat stronger, but still comparable to field strengths of a few kG expected from thin flux tube simulations that begin with $100$ kG at the base of the convection zone~\citep[see, e.g.,][]{Caligari:EmergingTubesPartI}. Thin flux tube simulations also predict a rise speed of between $0.5$ and $1$ km s$^{-1}$ at this depth. For this simulation, we chose to impose a rise speed of $1$ km s$^{-1}$.

The axis of symmetry of the torus is parallel to the $y$-axis of the Cartesian domain so that the resulting magnetic loop in the domain has its axis lying in the $x$-$z$ plane. At the bottom boundary, the gas pressure within the torus is given by $p_{\rm gas} = \langle p_{\rm bot}\rangle - B^2/8\pi$, where $\langle p_{\rm bot}\rangle = 2.44\times 10^{9}$ dyne cm$^{-2}$ is the mean gas pressure at the bottom boundary. The inflow specific entropy $s$ of the plasma in the torus is set to the same value as that of ambient convective upflows. This choice of the thermodynamic properties results in a relative density deficit of $\Delta \varrho/\langle \varrho_{\rm bot} \rangle \approx (\beta \gamma_1)^{-1} = 6\times 10^{-4}$ ($\gamma_1 = \frac{\partial \ln{p}}{\partial \ln{\varrho}}|_s $ is the first adiabatic exponent) at the tube axis. 

Beginning at $t=0$, the magnetic torus is kinematically advected into the domain with a constant rise speed of $1$ km s$^{-1}$ until the top half of the torus has traversed through the bottom boundary. Beyond this point in time (i.e., $t\ge5.9$ hrs), the velocity at the bottom boundary within the tube is set to zero. Outside of the tube, the bottom boundary condition allows for smooth inflows and outflows (see~\citeauthor{Voegler:MURaM} 2005 for details). Mass flows across the top boundary are allowed (upflows are unimpeded and downflows are damped). By virtue of the low densities near the top boundary, the mass flux across the top has a negligible effect on the mass content in the simulation domain. The magnetic field at the top boundary is matched to a potential field. Periodic boundary conditions apply at the side boundaries. Following~\citet*{Rempel:SunspotStructure}, we limit the strength of the Lorentz force in low-$\beta$ regions ($\beta < 0.05$) to limit the Alfv\'en speed to a maximum of $31$ km s$^{-1}$. This artificial limiting of the Lorentz force mainly acts within the central umbral regions of sunspots in mid-to-high photosphere, where local Alfv\'en speeds reach up to $10^3$ km s$^{-1}$ and would impose prohibitively small time-steps on the explicit scheme of the code.

Recently,~\citet*{MacTaggart:ToroidalFluxTubes} modeled the rise of buoyant toroidal magnetic flux tubes. Their simulation setup is different from ours in a number of ways. First of all, they used a plane-parallel background convection zone and atmospheric model to mimic the stratification in the Sun but do not include convective flows nor radiative transfer. Secondly, the toroidal flux tubes in their simulations were embedded as buoyant, stationary structures as an initial condition. An interesting result from their calculations is that the choice of a toroidal flux tube (as opposed to an originally horizontal tube) facilitated the emergence of the axis of the tube into the photosphere.~\citet*{MacTaggart:ToroidalFluxTubes} attributed this to the ability (in the toroidal tube case) of plasma to drain down the legs of of the tube to enhance buoyant at the tube apex. In section~\ref{sec:mass_discharge}, we showed that convective flows acting on the rising magnetic plasma have a similar effect.

\section{Simulation Results}

\subsection{Photospheric Emergence and Active Region Formation}

\begin{figure*}
\centering
\includegraphics[width=0.86\textwidth]{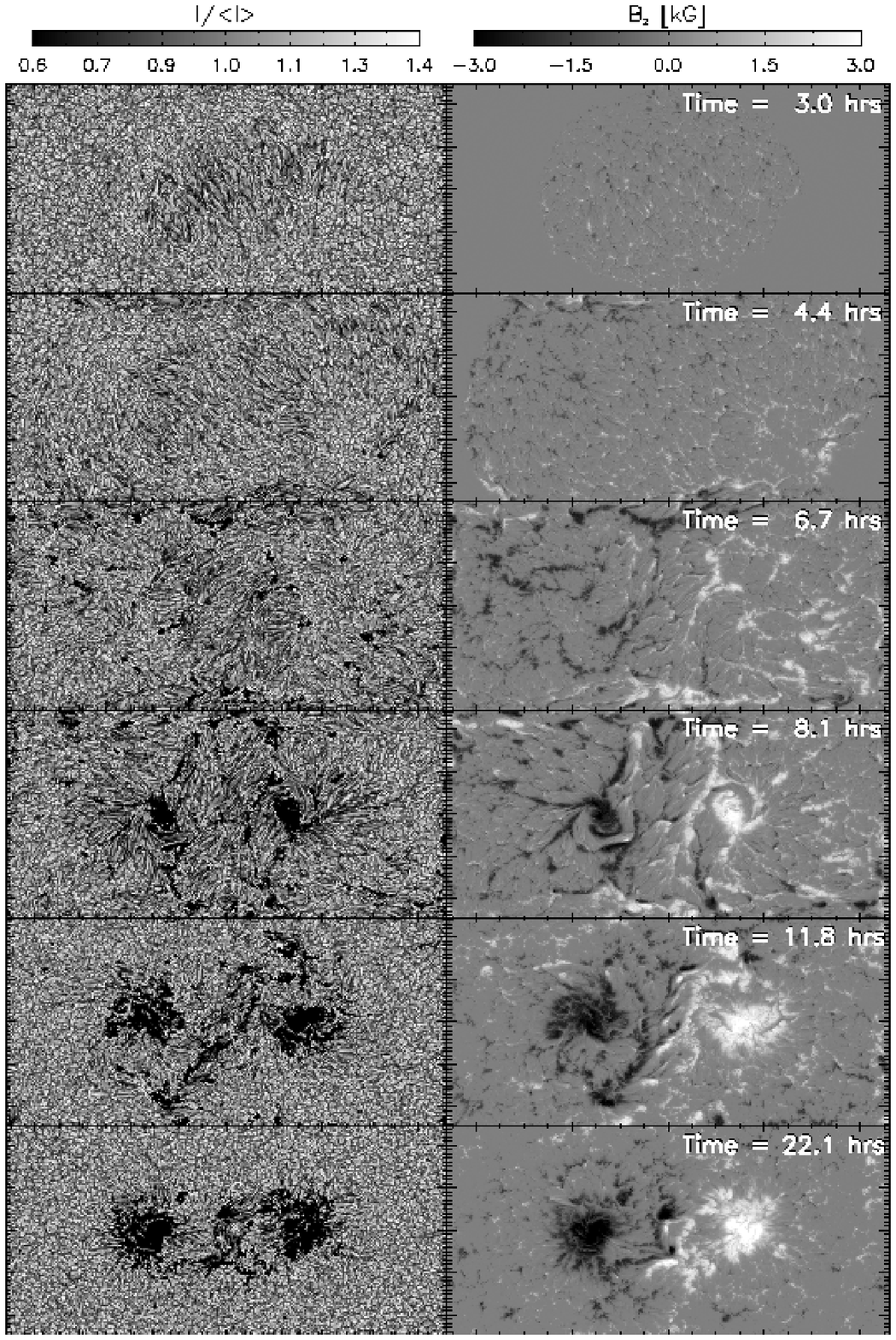}
\caption{Time sequence of continuum intensity images at $500$ nm (left) and synthetic longitudinal magnetograms (sampled at $\tau_{\rm Ross}=0.1$, right). The images show the full horizontal extent ($92 \times 49$ Mm) of the simulation domain.}
\label{fig1}
\end{figure*}

\begin{figure*}\centering
\includegraphics[width=0.9\textwidth]{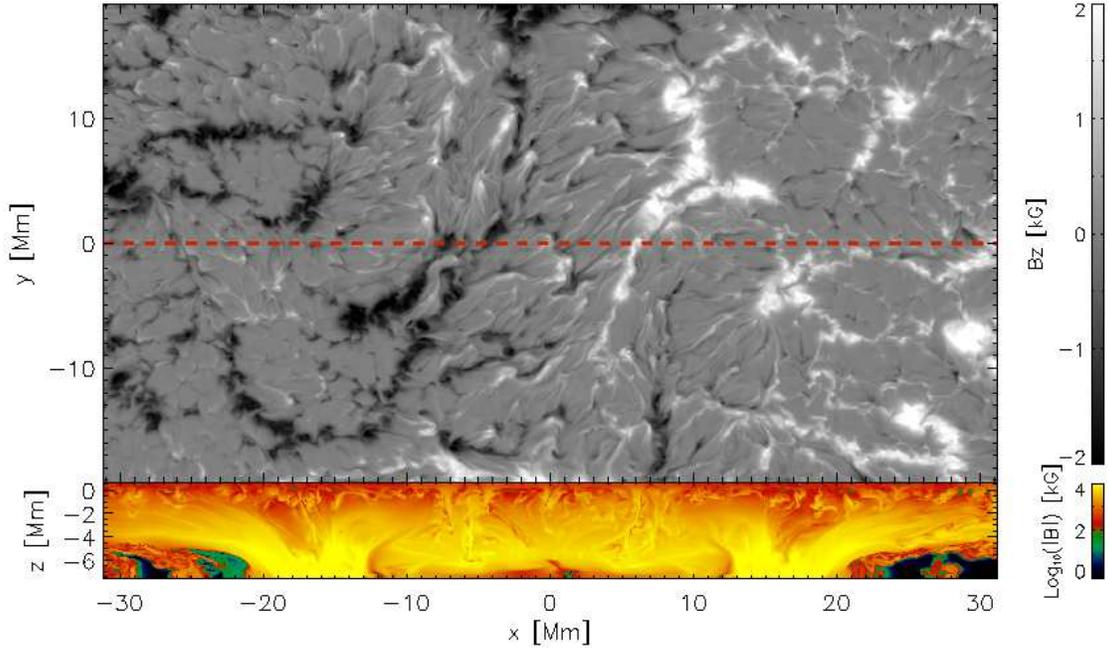}
\caption{The upper panel shows a synthetic magnetogram (sampled at $\tau_R = 0.1$) of the simulated emerging flux region at $t=6.6$ hrs. The lower panel shows the vertical cross-section of the magnetic field strength ($\log_{10}{B}$) along $y=0$ (delineated by red dashed line).}
\label{fig2}
\end{figure*}

\begin{figure*}\centering
\includegraphics[width=0.9\textwidth]{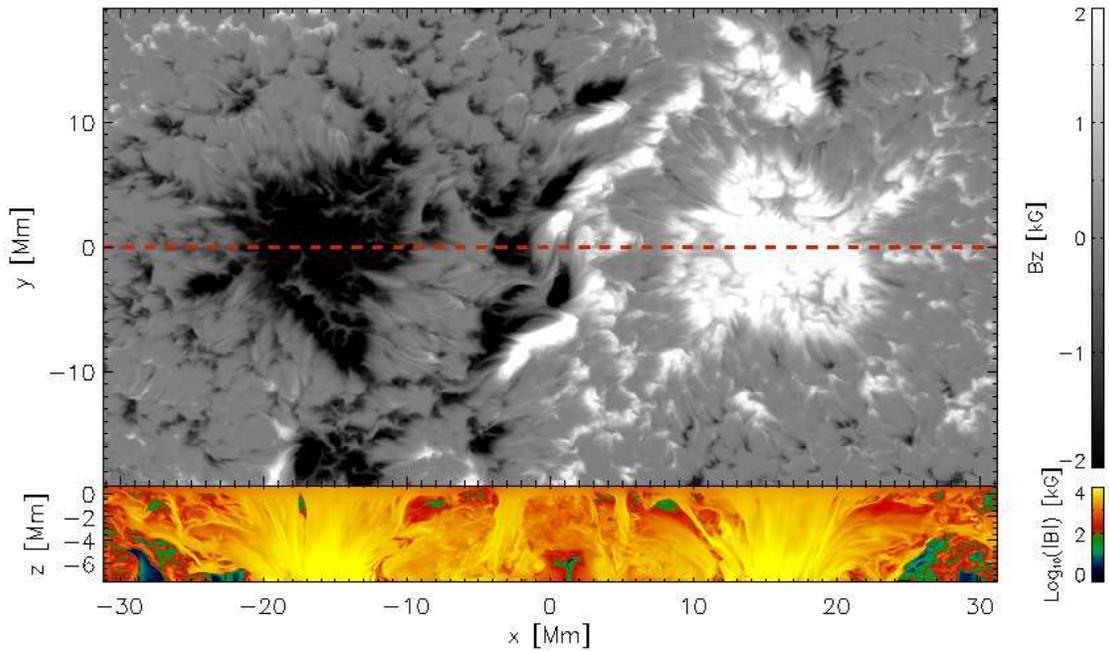}
\caption{Same as Fig~\ref{fig2} but at $t=15.3$ hrs. The two polarities of the active region at this time are much more compact.}\label{fig3}
\end{figure*}

Before proceeding to discuss the physics underlying the flux emergence and AR formation process, we describe how the emerging flux region evolves in time. Magnetic flux begins to emerge into the photosphere beginning at $t = 2$ hrs and emergence progresses for several hours. A time sequence of the grey intensity and maps of the vertical field (sampled at $\tau_{R}=0.1$) are shown in Fig.~\ref{fig1}. In the intensity map at $t=3$ hr, elongated granules begin to appear. This stretching of the convective cells is due to the horizontal expansion of the rising magnetic plasma. A more detailed discussion of the physics of horizontal expansion is given in section~\ref{sec:subsurface_rise}. Due to the subsurface horizontal expansion and undulation of the field lines by convective flows, the emerging flux region covers an extend surface area within which small-scale bipoles emerge with a systematic orientation~\citep[see also][]{Cheung:SolarSurfaceEmergingFluxRegions}. By $t=6.7$ hr, small pores begin to appear in multiple locations in the intensity map~\citep[c.f.][]{Stein:SolarFluxEmergenceSimulations}. A corresponding synthetic `magnetogram' ($B_z$ sampled at $\tau=0.1$) is shown in Fig.~\ref{fig2}. Two hours later, two spots have appeared near the horizontal positions coincident with where the subsurface footpoints of the initial semitorus are rooted. As time progresses, the intensity maps show the spots growing in size. A synthetic magnetogram at a later time shows two large, vertical concentrations of opposite polarity field at the locations of the spot (see Fig.~\ref{fig3}). Throughout the simulation, the two spots never attain fully developed penumbrae which is consistent with observations~\citep{Zwaan:AppearanceOfMagneticFlux,Zwaan:ElementsAndPatterns} since the two large magnetic concentrations have (over the course of the simulation) at most $4-5\times10^{21}$ Mx. Nevertheless, transient penumbral filaments, umbral dots and even light bridges are found in the synthetic intensity maps.

\subsection{Subsurface Rise and Expansion of the Magnetic Tube}\label{sec:subsurface_rise}

\begin{figure*}
\centering
\subfigure[]{\includegraphics[width=0.475\textwidth]{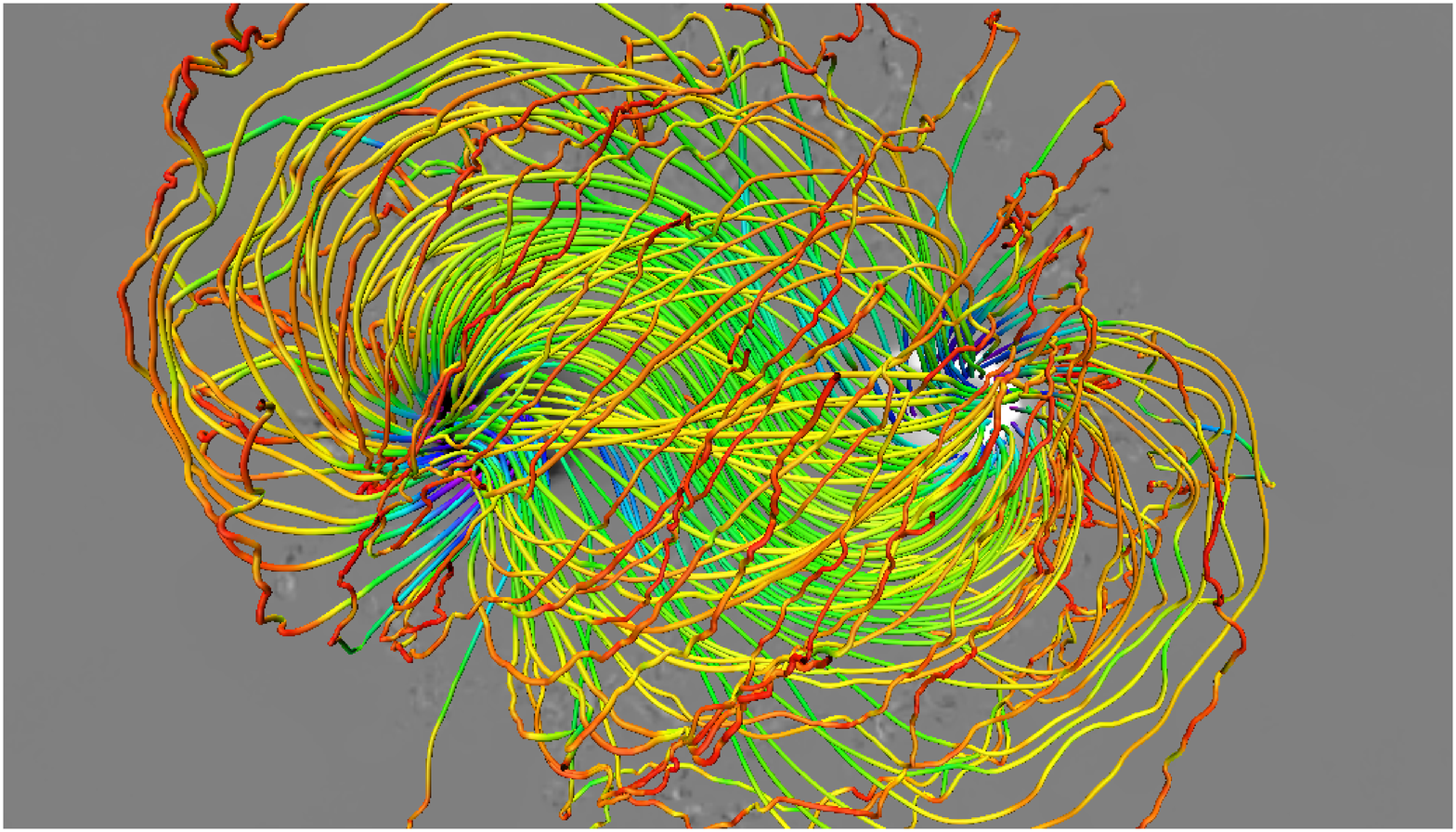}}  \hspace{0.01\textwidth}
\subfigure[]{\includegraphics[width=0.475\textwidth]{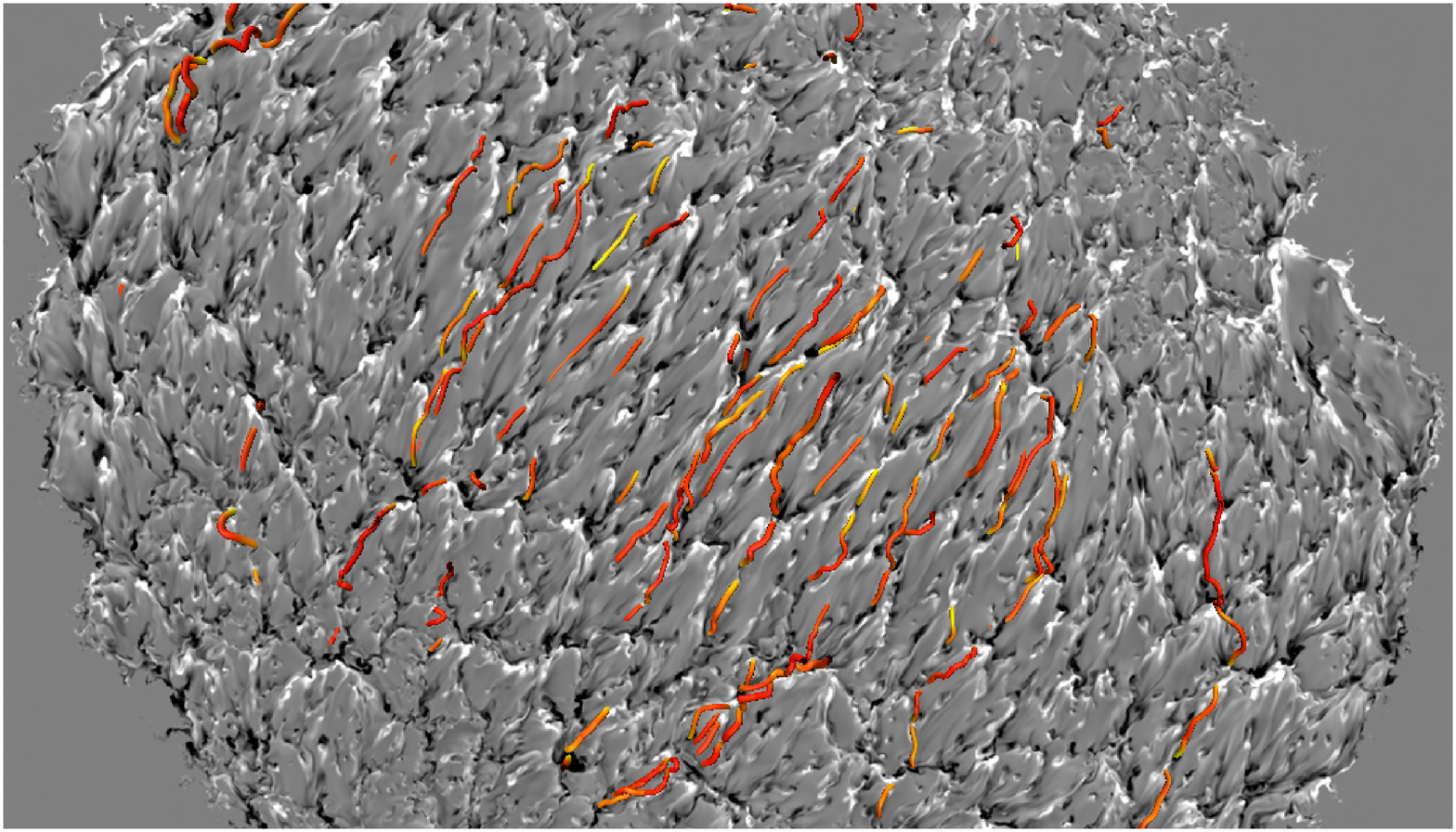}}
\caption{Magnetic structure of the (semi)torus as it erupts onto the solar surface ($t = 4$ hrs). The same set of field lines are shown in both panels. The left panel shows the distribution of the vertical component of the magnetic field (scaled between $\pm 17$ kG) at $z=-7.5$ Mm. At this depth the opposite polarities are in the form of two coherent roots. The panel on the right shows the corresponding magnetic map at $z=0$ (scaled between $\pm 2$ kG), which highlights the serpentine nature of the magnetic field lines near the surface as a consequence of the interaction between the emerging field and the granular convective flows.}\label{fig4}
\end{figure*}

Figure~\ref{fig4} shows the structure of the torus as it begins to erupt from the convection zone into the overlying photosphere. The subsurface roots of the torus at the bottom of the computational domain are shown in the magnetic map at a depth of 7.5 Mm (left panel). Field lines traced from the opposite polarity roots diverge upwards to covered an extended horizontal region. The right panel shows the associated emerging flux region at the surface, where the magnetic field lines have been undulated by their interaction with the convective downflows as is consistent with observations~\citep[see e.g.][]{Pariat:ResistiveEmergence,Cheung:FluxEmergenceInGranularConvection,Centeno:SmallscaleFluxEmergence,MartinezGonzalez:LowLyingLoops,MartinezGonzalez:SmallScaleLoops,Cheung:SolarSurfaceEmergingFluxRegions,Ishikawa:3DView}.

\begin{figure*}
\centering
\includegraphics[width=0.75\textwidth]{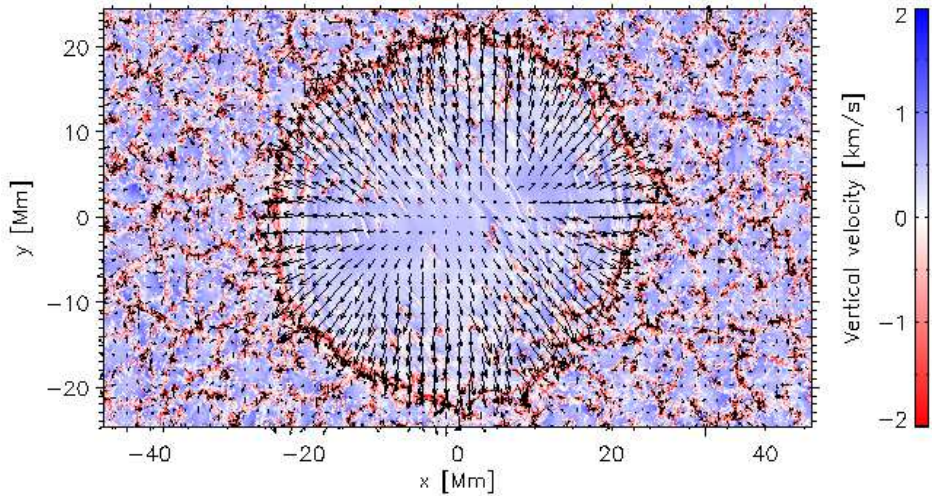}
\caption{Velocity field below the surface ($z=-2$ Mm) as the emerging magnetic field is traversing this layer ($t=2.8$ hrs). Red and blue colors indicate down- and upflows, respectively, and the arrows indicate the horizontal velocity. The outflows at the periphery of the emerging flux region reaches speeds of up to $\sim 4$ km s$^{-1}$ whereas upflow velocities are closer to  $\sim 0.5$ km s$^{-1}$.}\label{fig5}
\end{figure*}

The number of pressure scale-heights spanned between $z=-7.5$ Mm and $z=0$ (i.e. the photospheric base) is $N(H_p) = 10$. To reach pressure equilibrium with the surroundings, the rising magnetic structure expands strongly in the horizontal directions. This is an important stage in the evolution for reasons of hydrostatic balance and mass conservation. Since the rise time of the flux tube is much longer than both the free-fall and sound-crossing times (both $\sim 10^2$ s) across the layers of the convection zone captured in this simulation, its rise does not significantly alter the mean stratification. On the other hand, the injection of mass from the rise of the half-torus is equal to $40\%$ of the original mass in the entire domain, which is roughly equal to the~\emph{total} mass content of the top $6$ Mm of the convection zone (as captured in the computational domain). Thus it is not surprising that the rising magnetic plasma displaces a substantial fraction of the original mass in the domain and fills the near-surface layers during the first stage of the active region formation process. As a consequence, the top of the magnetic torus takes on a flattened, pancake-like structure in near-surface layers (see Fig.~\ref{fig5}).~{A similar behavior is also reported in the flux emergence simulations by~\citet{Toriumi:TwoStepEmergence}}.

Given the field strength of the original magnetic torus at a depth of $7.5$ Mm, what strengths can we expect for the field emerging into the photosphere? This question can be addressed by considering a number of different simplified scenarios. For a purely horizontal magnetic flux tube expanding in directions transverse to its axis, conservation of mass and conservation of magnetic flux give the linear scaling relation $B/\varrho={\rm constant}$. A different scaling relation can be derived by considering how an upwelling in the stratified convection zone modifies horizontal field threading the rising plasma. Let the initial horizontal field be $\vec{\bf B} = B\hat{x}$ and for simplicity assume that the upflow is axisymmetric about the (vertical) $z$-axis and centered at the origin (the absolute $z$ coordinate is unimportant). Let the rates of expansion in the horizontal and vertical directions be  $\alpha =
\partial v_x/\partial x = \partial v_y/ \partial y$ and $\varepsilon \alpha =
\partial v_z / \partial z$ respectively. Here $\varepsilon$ is a measure of the flow anisotropy ($\varepsilon=1$ for isotropic expansion). From the ideal Induction Equation, the rate of change of the magnetic field attached to a Lagrangian fluid element is 
\begin{equation}
\frac{D\vec{\bf B}}{Dt} = -(\nabla \cdot \vec{\bf v})\vec{\bf B} +
(\vec{\bf B}\cdot \nabla) \vec{\bf v}, 
\end{equation}
\noindent which in this case reduces to 
\begin{equation}
\frac{D\ln{B}}{Dt} = -(1 + \varepsilon)\alpha.\label{eq:induction_special}
\end{equation} 
\noindent The corresponding Lagrangian form of the continuity equation is 
\begin{eqnarray}
\frac{D \ln{\varrho}}{Dt} &=& -\nabla \cdot {\vec{\bf v}} \\
&=& -(2 + \varepsilon)\alpha.\label{eq:continuity_special}
\end{eqnarray}
\noindent Combining equations (\ref{eq:induction_special}) and (\ref{eq:continuity_special}), we obtain the scaling relation 
\begin{equation}
B \propto \varrho^{(1+\varepsilon)/(2+\varepsilon)}.
\end{equation}
\noindent For isotropic expansion ($\varepsilon = 1$),  $B\propto
\varrho^{2/3}$. In the case where the horizontal expansion rate is much
larger than that of the vertical expansion rate we have $\varepsilon \ll 1$, so that $B\propto
\sqrt{\varrho}$. 

For a field strength of $10$ kG at $\varrho=4\times 10^{-4}$ g cm$^{-3}$ (density at $z=-7.5$ Mm), the linear scaling relation gives, for $\varrho_{\rm ph}\sim 4\times 10^{-7}$g cm$^{-3}$, a photospheric field strength of $\sim 10$ G. This is evidently too weak compared to what is found in the simulation and to what is observed, both of which give emerging horizontal field strengths on the order of a few hundred G~\citep{Lites:EmergingFieldsVector,Kubo:EmergingFluxRegion}. 

\begin{figure}
\centering
\includegraphics[width=0.49\textwidth]{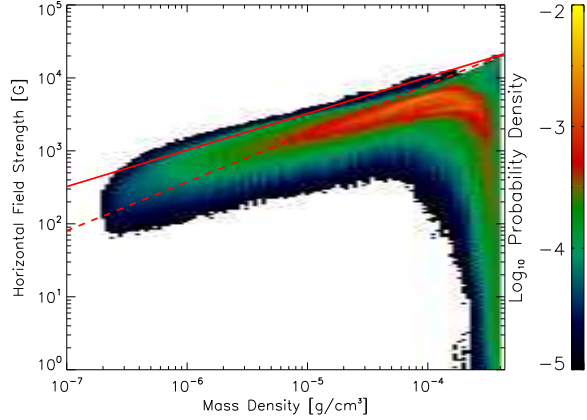}
\caption{Joint probability density function of the horizontal field strength. The values were sampled at the $x=0$ vertical plane between $t=1$ and $t=8$ hours. The solid and dashed lines respectively show the scaling relations $B \propto \varrho^{1/2}$ and $B \propto \varrho^{2/3}$ with $B_0=21$ kG and $\varrho_0=4.2\times10^{-4}$ g cm$^{-3}$. }
\label{fig:rhoB_scaling}
\end{figure}

The second scaling relation $B\propto \varrho^{\frac{1+\varepsilon}{2+\varepsilon}}$ yields more consistent values for emerging horizontal field strengths. For $\varepsilon = 0$ (zero expansion in the vertical direction) and $\varepsilon=1$ (isotropic expansion), a density drop of $10^3$ yields horizontal photospheric field strengths of $B=100$ G and $B=300$ G, respectively. Figure~\ref{fig:rhoB_scaling} shows a joint probability density function (JPDF) of the horizontal field strength versus mass density sampled at the mid-plane $x=0$ between $t=1$ and $t=8$ hours. Points with low values of specific entropy ($s <  9.75\times10^8$) have been excluded since they correspond to plasma that has undergone radiative cooling at the surface. The maximum magnetic field strength of $21$ kG corresponds to the field strength at the axis of the torus that is introduced through the bottom boundary. The solid and dashed lines correspond to the scaling relations $B\propto \sqrt{\varrho}$ and $B\propto \varrho^{2/3}$ respectively. The latter scaling relation is clearly too steep and underestimates the field strengths at photospheric densities ($\varrho_{\rm ph}\approx 4\times 10^{-7}$ g cm$^{-3}$). The square-root scaling relation provides a much better match to the JPDF. This result is consistent with the fact that the plasma experiences predominantly horizontal expansion ($\varepsilon $ close to zero) during its rise to the surface (see Fig.~\ref{fig5}).

\begin{figure*}
\centering
\includegraphics[width=\textwidth]{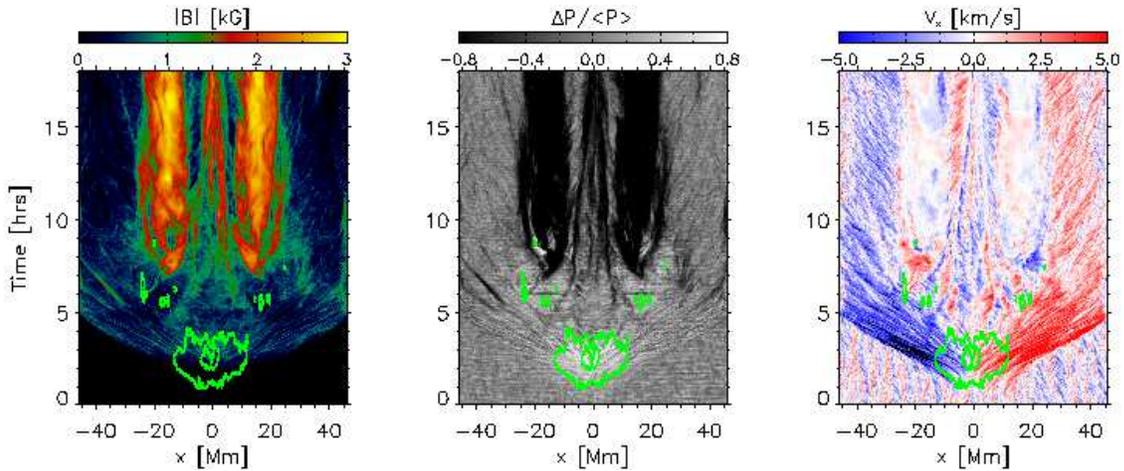}
\caption{The emerging flux tube leads to a enhancement of pressure which drives outflow away from the emergence site. The three panels in this figure show, respectively, the time evolution of the surface (photospheric base, $z=0$) magnetic field strength, relative gas pressure perturbation, and $x-$component of velocity averaged over the band $y\in[-2,2]$ Mm. In all three panels, the green contours indicate enhancements of the gas pressure (relative to $\langle P_{\rm gas}\rangle = 9\times10^4$ dyne cm$^{-2}$) by $25\%$ and $50\%$ (for the purpose of having fewer contours, the pressure values have been smoothed in time with a Gaussian filter with $\sigma=30$ min).}
\label{fig:outflows}
\end{figure*}

Figure~\ref{fig:outflows} shows how the arrival of buoyant, magnetic plasma at the photosphere leads to a transient pressure excess which drives diverging horizontal flows about the flux emergence site. The coincidence (spatially and temporally) of this pressure enhancement with the onset of large-scale diverging flows (beginning at about $t=1-2$ hrs) is evidence that the flows are driven by the associated horizontal pressure gradient. $\Delta P/\langle P \rangle$ is of the order $0.1-0.5$, so that the horizontal flows have Mach numbers of similar amplitude. After the initial emergence phase ($t > 6$ hr), the divergent horizontal flows at the periphery of the magnetic complex persist, albeit with lower amplitude ($|v_x|\sim 1-3$ km s$^{-1}$). This raises the question of how magnetic flux is still capable of accumulating in the two spots, which will be addressed in section~\ref{sec:flux_growth}.

\subsection{Discharge of Mass from the Rising Magnetic Structure}\label{sec:mass_discharge}

Given that the near-surface field is rather dispersed, the next stage of active region evolution is the coalescence of the dispersed field into coherent concentrations. In order for the fragments to collect together, excess mass must be unloaded from the emerged field lines. To illustrate the amount of mass that is eventually removed, we compare the mass carried within the original half-torus and the mass remaining within the two opposite polarity concentrations at $t=8$ hrs. Let $\overline{\bf B}(r,z)$ and $\overline{\varrho}(r,z)$ be azimuthal averages of one such flux concentration about its vertical axis. Using these, we calculated surfaces of constant magnetic flux and the corresponding mass content contained within the flux surfaces. The mass contained within a volume defined by the flux surface $\Phi(r,z)=4\times 10^{21}$ Mx (summed over both spots) at $t=8$ hrs is only $12\%$ of the mass contained in a sub-volume of the original half-torus with the same flux content. Thus most of the original injected mass has somehow escaped to the surroundings. 

{
A number of physical mechanisms may be invoked to explain the discharge of mass from the original magnetic structure. First of all, magnetic diffusion of the torus allows mass transfer across magnetic field lines. However, this would result in an increase of mass in the semitorus rather than a decrease and thus can be ruled out as the underlying mechanism for mass discharge. As a second possibility, we consider the mass flux through the top and bottom boundaries. By virtue of the low densities at the top boundary, the vertical mass flux there is negligible for the mass budget of the semitorus. At the bottom boundary, vertical (and horizontal) velocities within the torus ($ \omega \le \sqrt{2}a$) are set to zero after the initial prescribed rise. Downflows are permitted at other magnetic footpoints at the bottom boundary but since the majority of the magnetic flux at that layer remains within the torus, mass flux through the bottom boundary can also be ruled out as the main discharge mechanism. A third possibility is the removal of mass via outflows associated with the horizontal expansion (see section~\ref{sec:subsurface_rise}). While it is true that outflows carry mass away from the center of the emerging region, these flows also advect magnetic field with them and cause a weakening of the field. So outflows alone are insufficient to explain how mass is removed from the field lines. }

\begin{figure*}
\centering
\subfigure[]{\includegraphics[width=0.42\textwidth]{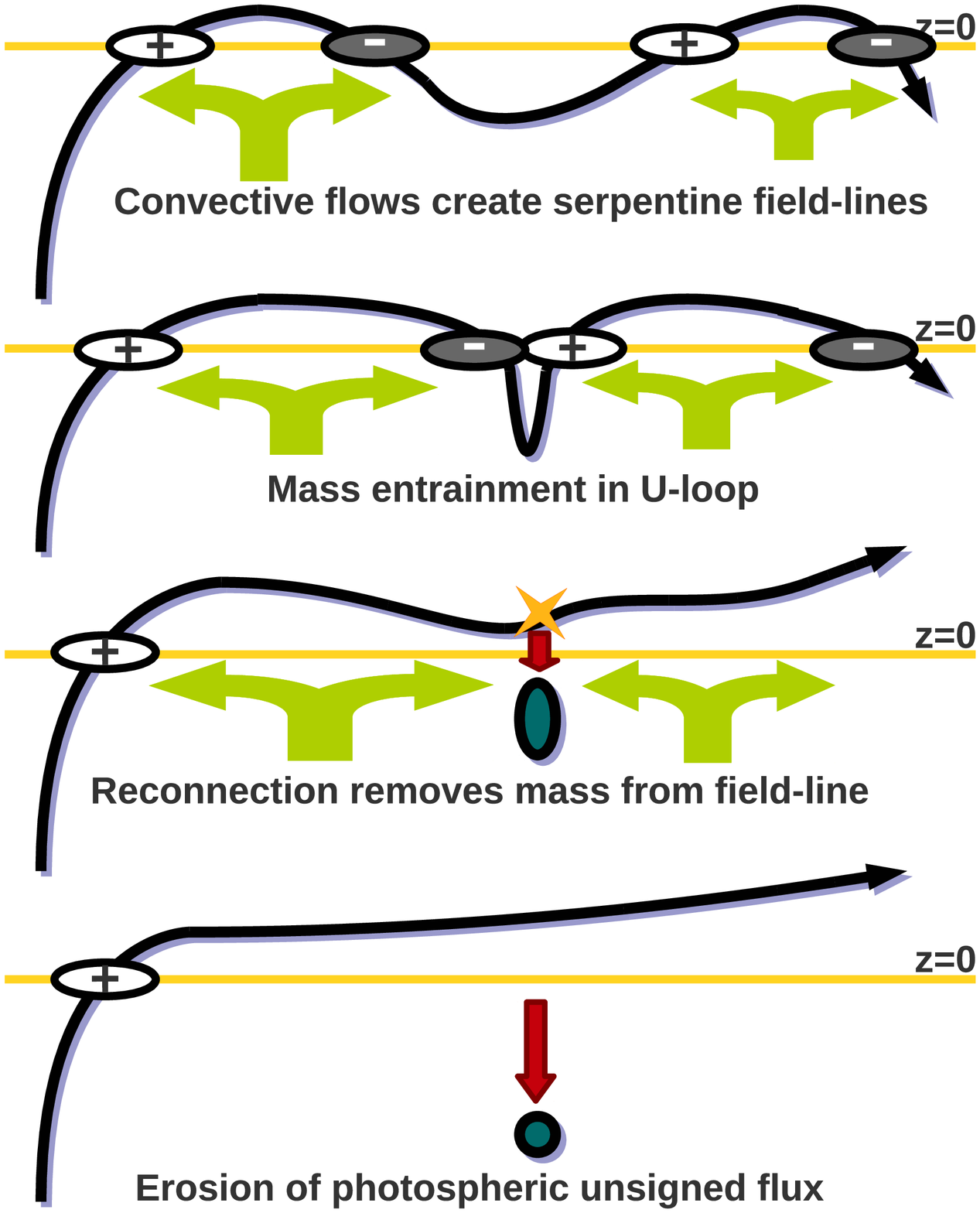}}  \hspace{0.01\textwidth}
\subfigure[]{\includegraphics[width=0.47\textwidth]{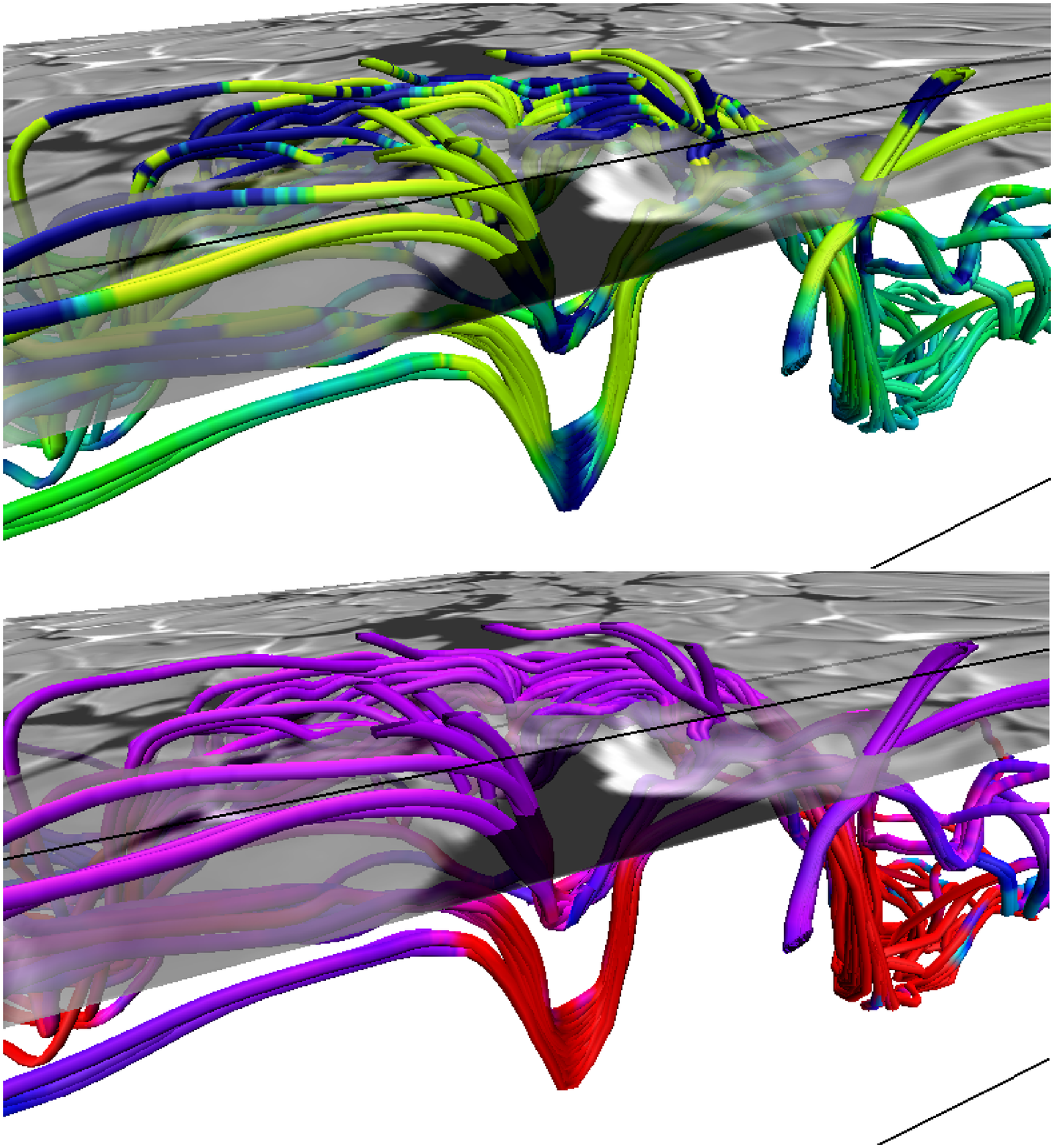}}
\caption{Mechanism for removal of mass and unsigned flux from the surface. (a) Schematic representation of how mass is removed from emerging magnetic field lines in a 2D scenario. In addition to undulating field lines, convective flows expel emerged flux (indicated by ovals labeled with positive and negative signs) from granular upflows. (b) 3D rendering of near-surface field lines in the simulated emerging flux region. Field lines in the upper panel are colored in accordance with the local density perturbation (about horizontal mean) with dark blue indicating density enhancement. Field lines in the lower panel are colored according to the vertical component of the momentum with red indicating downflowing material.}\label{fig:mass_removal}
\end{figure*}

{The responsible mass removal mechanism is illustrated} in Fig.~\ref{fig:mass_removal}. The left panel of this figure shows a schematic representation of convective flows undulating a field line which has one end anchored deep below the photosphere. As already reported in~\citet{Cheung:SolarSurfaceEmergingFluxRegions}, expulsion of flux from the granular centers lead to the encounter and cancellation of opposite polarity field. Such opposite polarity pairs are connected below the surface in the form of U-loops~\citep[see also][]{Stein:SolarFluxEmergenceSimulations}. The first consequence of this magnetic reconnection between the opposite polarities is the creation of an O-loop, which discharges mass from the original field line.~{This is akin to the process suggested by~\citet{Parker:OmegaPumping} for the discharge of mass from and intensification of magnetic fields at the bottom of the convection zone}. Based on observations of emerging flux regions with Spectro-Polarimeter instrument~\citep{Lites:HinodeSP} of the Solar Optical Telescope~\citep{Tsuneta:SOT} onboard the Hinode satellite~\citep{Kosugi:Hinode},~\citet{Lites:SpaceScienceReviews} put forward this mechanism to explain how mass is discharged from magnetic field lines in emerging flux regions. In the simulation, mass is subducted in a similar fashion though in the 3D case plasmoids take the place of O-loops.

The amount of vertical unsigned flux at the surface may erode depending on the height at which reconnection occurs. Specifically, if reconnection happens exactly at the surface, the unsigned flux will immediately decrease. If reconnection occurs above $\tau=1$, the unsigned flux will decrease only when resulting O-loops (plasmoids) submerge. This second scenario has indeed been reported by~\citet*{Iida:CancellationSite}. If reconnection occurs below the surface, the surface unsigned flux will decrease if the remnant U-loop (counterpart to the O-loop after reconnection) rises above the surface. This is less likely since the U-loop will still be anchored to downflows. We find in our simulation that reconnection occurs both in the photosphere and in the convection zone. However, the unsigned flux (at any one horizontal level) erodes due to retraction of inverse O-loops (plasmoids). The process described has been numerically modeled by~\citet*[][in 2D]{Isobe:EllermanBombs} and by~\citet*[][in 3D]{Archontis:EllermanBombs} for explaining the origin of Ellerman bombs resulting from magnetic reconnection occurring in emerging flux regions~\citep{Pariat:ResistiveEmergence,Pariat:SerpentineFluxTubes}.

\begin{figure}
\centering
\includegraphics[width=0.49\textwidth]{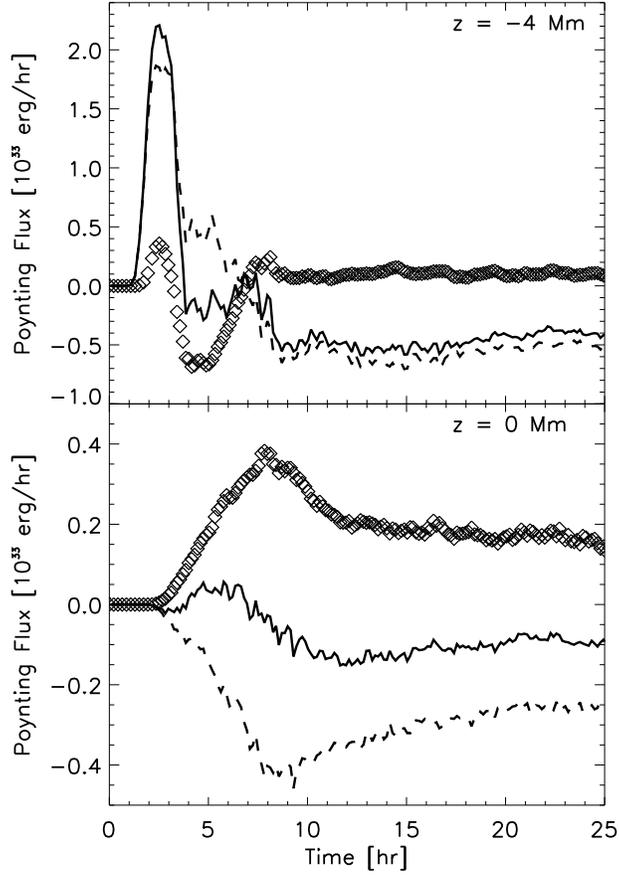}
\caption{Poynting fluxes of magnetic energy through horizontal planes at $z=0$ and $z=-4$ Mm. In both plots, the dashed curves and diamonds indicate the contributions by the first (bodily emergence or submergence of horizontal fields) and second (shearing by horizontal flows) terms in Eq.~(\ref{eqn:poynting}), respectively. The solid curves indicate the sum of the two contributions.}
\label{fig:poynting}
\end{figure}

The right panel of Fig.~\ref{fig:mass_removal} shows an example of a collection of sinking U-loops in the simulation. The downward transport of such U-loops, which allows the discharge of mass from the rising magnetic field, is related to the phenomenon known as~\emph{turbulent pumping}~\citep{Tobias:TurbulentPumping,Brummell:TurbulentPumping}. Turbulent pumping manifests itself in terms of vertical component the Poynting flux of magnetic energy, which for ideal MHD is 
\begin{equation}
S_z = \frac{1}{4\pi}\int v_z (B_x^2+B_y^2) - B_z (B_x v_x + B_y v_y) dx dy,\label{eqn:poynting}
\end{equation}
\noindent where the first term in the integrand represents the bodily ascent (or descent) of horizontal fields and the second term represents the Poynting flux due to horizontal flows. Figure~\ref{fig:poynting} shows plots of the Poynting flux through two horizontal planes ($z=0$ and $z=-4$ Mm) as functions of time. The individual contributions from vertical and horizontal flows are also plotted. The traversal of the rising magnetic structure through the $z=-4$ Mm plane appears as a positive hump in both the emergence term and the combined Poynting flux between $t=1.5$ and $t=4$ hr. Thereafter the effect of turbulent pumping takes over and the net Poynting flux due to vertical flows is negative. At the surface ($z=0$ Mm), the convective downflows are much stronger (in terms of Mach number, they approach $0.1$) and the effect of turbulent pumping on the Poynting flux is pronounced throughout the emergence and post-emergence phase. It should be noted that although the Poynting flux (a flux of energy) has a negative sign, it does not necessarily imply that net magnetic flux itself is migrating downwards.

\subsection{Growth of Magnetic Flux Content in the Developing Spots}\label{sec:flux_growth}
As already discussed in the previous section, mean horizontal outflows persist through the simulated duration of AR formation. Given such mean flows, how does the magnetic flux at the photosphere migrate inwards to yield coherent spots? To examine this, the magnetic and velocity fields (for a fixed time) were decomposed into the sum of azimuthal averages and corresponding fluctuating components, namely $\vec{\bf B}(r,\theta,z)= \overline{\bf B}(r,z) + \vec{\bf B}'(r,\theta,z)$ etc, where the bar and prime denote an azimuthal average and the fluctuation about the average, respectively. For ideal MHD, the mean field induction equation is 
\begin{equation}
\frac{\partial \overline{\bf B}}{\partial t} = \nabla \times (\overline{\bf v}\times\overline{\bf B}) + \nabla \times \vec{\bf \mathcal{E}},
\end{equation}
\noindent where $ \vec{\bf \mathcal{E}} = \overline{\bf \vec{\bf v}'\times\vec{\bf B}'} $ is the (azimuthally-averaged) mean-field electromotive force resulting from correlations between the magnetic field and velocity field fluctuations~\citep{KrauseRaedler:MeanField,Raedler:MeanField}.  To examine how (vertical) magnetic flux accumulates in the growing spots, consider a circular area $\mathcal{C}$ of fixed radius $R$, at height $z=0$ and centered at $r=0$. The time rate of change of the total magnetic flux crossing this circular area is 
\begin{eqnarray}
\dot{\Phi}_\mathcal{C} & = & \int_\mathcal{C} \frac{\partial \overline{\bf B}}{\partial t} \cdot \vec{\bf dS} \\
& = & \int_\mathcal{C} \nabla \times \left(\overline{\bf v} \times \overline{\bf B}  +  \vec{\bf \mathcal{E}}\right) \cdot\vec{\bf dS}, \\
& = & \oint_{\partial \mathcal{C}} \left(\overline{\bf v} \times \overline{\bf B} +  \vec{\bf \mathcal{E}} \right)\cdot \vec{\bf dl}, \\
& = & \dot{\Phi}_{\rm m} + \dot{\Phi}_{\rm f}\label{eq:mean_induction}, 
\end{eqnarray}
\noindent where 
\begin{eqnarray}
\dot{\Phi}_{\rm m} &=& 2\pi R[\overline{\bf v} \times \overline{\bf B}]_\theta, \\
&=& 2\pi R(\overline{v}_z\overline{B}_r - \overline{B}_z\overline{v}_r),\\
\dot{\Phi}_{\rm f} &=& 2\pi R\mathcal{E}_\theta, \\
&=& 2\pi R\overline{v_z'B_r' - B_z'v_r'} .
\end{eqnarray}
\noindent Equation~(\ref{eq:mean_induction}) indicates that the magnetic flux enclosed within $\mathcal{C}$ changes as a result of (1) advection of the mean magnetic field $\overline{\bf B}$ by the mean flow $\overline{v}$ (first term on the r.h.s.), and (2) correlations between their fluctuating components.

\begin{figure}
\includegraphics[width=0.48\textwidth]{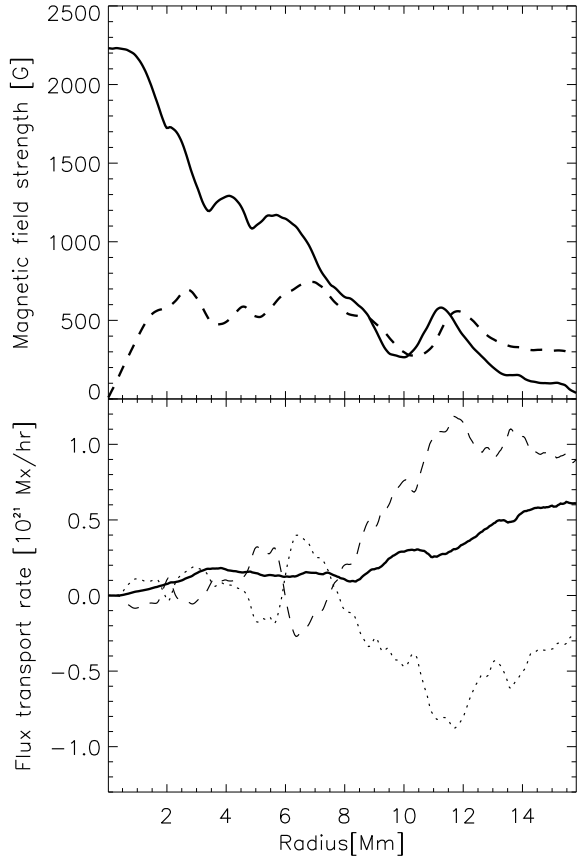}
\caption{The top panel shows the azimuthally- and temporally-averaged (between $t=9.3$ and $t=10$ hr, sampled at $z=0$) magnetic field strengths (solid and dashed lines indicate the vertical and radial components, respectively) as functions of the radial distance from the axis of the positive-polarity spot in the simulation. The lower panels shows the corresponding quantities as defined in Eq. (\ref{eq:mean_induction}): $\dot{\Phi}_{\rm m}$ (dotted), $\dot{\Phi}_{\rm f}$ (dashed) and their sum (solid). Positive values of $\dot{\Phi}$ indicate an increase of flux (within the area $r<R$) with time.}\label{fig:flux_transport}
\end{figure}

Figure~\ref{fig:flux_transport} shows plots of the mean magnetic field and of the flux transport rates defined in Eq. (\ref{eq:mean_induction}). To smooth out fluctuations in time, the azimuthal means and azimuthally-fluctuating quantities were also averaged between $t=9.3$ and $t=10$ hr in time. This remains consistent with the derivation above since the operations of averaging temporally and azimuthally commute. At this stage of the simulation, the mean magnetic field within a radius of $8$ Mm has already reached kG values and we focus our attention on how the weaker field at larger radii is transported. Inspection of the mean velocity field reveals that the contribution from the radially directed outflow ($\overline{v}_r>0$) dominates such that magnetic flux is transported~\emph{away} from the spot. However, it is found that correlations between the velocity and magnetic fields (i.e. $\vec{\bf \mathcal{E}}$) provide a means for the inward migration of vertical field. The amplitude of $\dot{\Phi}_{\rm f}$ is in fact larger than that of $\dot{\Phi}_{\rm m}$ and the net result is an accumulation of flux in the spot.

{The streaming of opposite polarity (vertical) field in opposite horizontal directions~\citep{StrousZwaan:HorizontalDynamics,StrousZwaan:SmallScaleStructure,Bernasconi:EmergingFlux,Cheung:SolarSurfaceEmergingFluxRegions} provides the systematic correlation needed for accumulation of magnetic flux in the spot. This begs the question of what is the underlying driver of the counter-streaming motion. The left panel of Figure~\ref{fig:mass_removal} shows how the granular expulsion~\citep{Weiss:FluxExpulsion,Hurlburt:Magnetoconvection} of magnetic flux from serpentine field lines leads to a migration of positive polarity in one horizontal direction and negative polarity field in the opposite direction (i.e. a non-zero $\vec{\bf \mathcal{E}}$). The presence of sea serpents, however, is an insufficient condition for flux accumulation in spots. For instance, consider a scenario wherein uniform magnetic field (with a net horizontal component) threads a series of granules. Although $\vec{\bf \mathcal{E}}$ is non-zero, the translational symmetry of the setup causes the loop integral  $\oint_{\partial \mathcal{C}}\vec{\bf \mathcal{E}}\cdot \vec{\bf dl} $ to vanish so that there is no~\emph{net} change of flux.}

\begin{figure}
\includegraphics[width=0.48\textwidth]{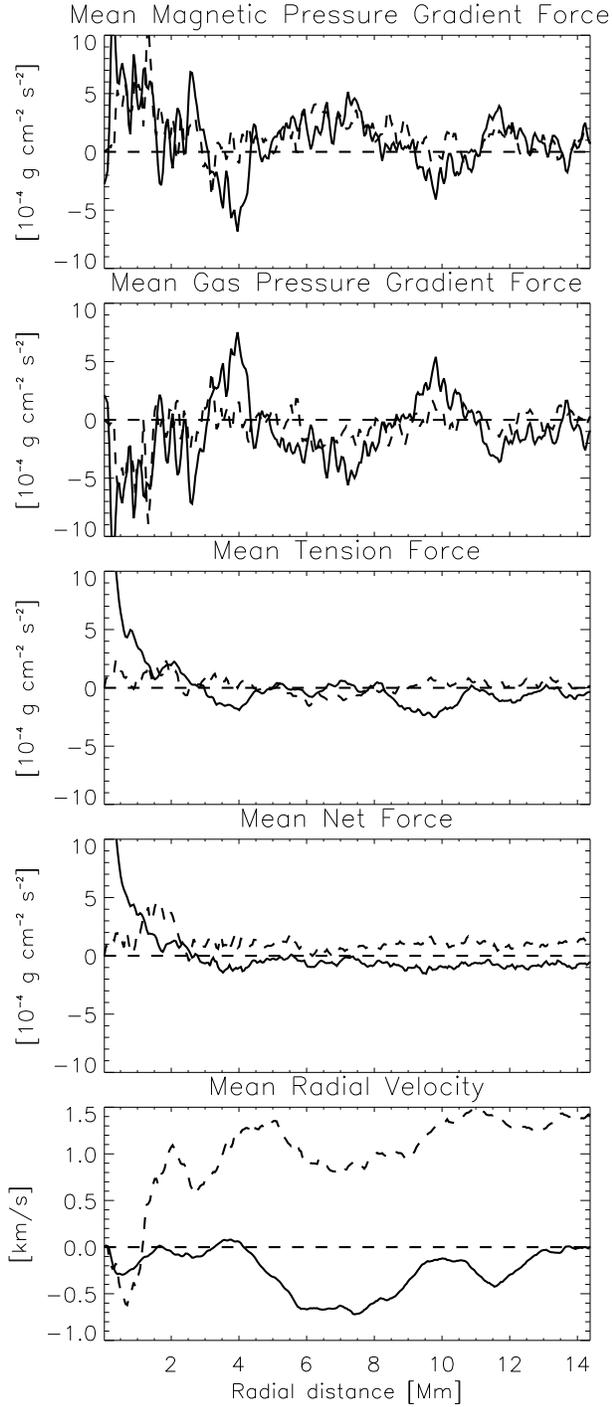}
\caption{The panels above show the azimuthally-, temporally- and spatially- (over a height-range of $540$ km about $z=0$) averaged profiles of the radial components of the magnetic pressure gradient force ($-\nabla B^2/8\pi$), gas pressure gradient force ($-\nabla p_{\rm gas}$), magnetic tension ($\vec{\bf B}\cdot\nabla \vec{\bf B}/4\pi$), net force and velocity ($\vec{\bf v}$) for gas with positive polarity field ($B_z > 0$, thick solid lines) and negative polarity field ($B_z<0$, thick dashed lines). The developing spot centered at $r=0$ has positive polarity.}\label{fig:ave_lorentz}
\end{figure}

{ The missing ingredient is some underlying large-scale structure which remains coherent over time-scales much greater than the granulation turnover times at the surface. Here it is provided by the coherent subsurface roots of the nascent active region (which is illustrated by the anchored field line in the left panel of Fig.~\ref{fig:mass_removal}). The upper four panels of Fig.~\ref{fig:ave_lorentz} show radial profiles of the azimuthally-, temporally- and spatially- (over a height of $540$ km about $z=0$) averaged forces about the developing spot. For both positive and negative polarities, the gas and magnetic pressure gradient forces roughly (but not exactly) balance each other. The profile for the magnetic tension force ($\vec{\bf B}\cdot \nabla \vec{\bf B}/4\pi$) shows a tendency to accelerate positive (negative) polarities inward (outward). This trend is even clearer when one looks at the net force (-$\nabla(p_{\rm gas}+B^2/8\pi)+\vec{\bf B}\cdot \nabla \vec{\bf B}/4\pi$) and is consistent with the  counter-streaming behavior of the opposite polarities (see bottom panel of Fig.~\ref{fig:ave_lorentz}). From this analysis, we conclude that the organization of the surface flux fragments into the two spots results from the presence of the coherent surface roots which influence the surface dynamics via the Lorentz force (especially the tension force).}

{This result closely resembles the~\emph{tethered-balloon} model of~\citet{Spruit:ClusterModel}. In this model, emerging magnetic loops consist of buoyant gas in the crests and neutrally-buoyant or anti-buoyant gas at footpoints anchored below the photosphere. Like tethered balloons, the equilibrium state is reached when buoyant elements hover vertically above their footpoints. The evolution of a field-line segment from the oblique to the vertical is mediated by the tension force. In our case, the interaction of granular convective flows with emerging magnetic fields, resulting in the undulation of field lines, flux expulsion to intergranular lanes and subsequent convective intensification~\citep{Cheung:SolarSurfaceEmergingFluxRegions,Danilovic:FieldIntensification} add a layer of complexity to the problem. Nevertheless, the present analysis demonstrates that Spruit's model is qualitatively compatible with our simulation results. }

\subsection{Light Bridge Formation and Disappearance}
\begin{figure*}\centering
\includegraphics[width=\textwidth]{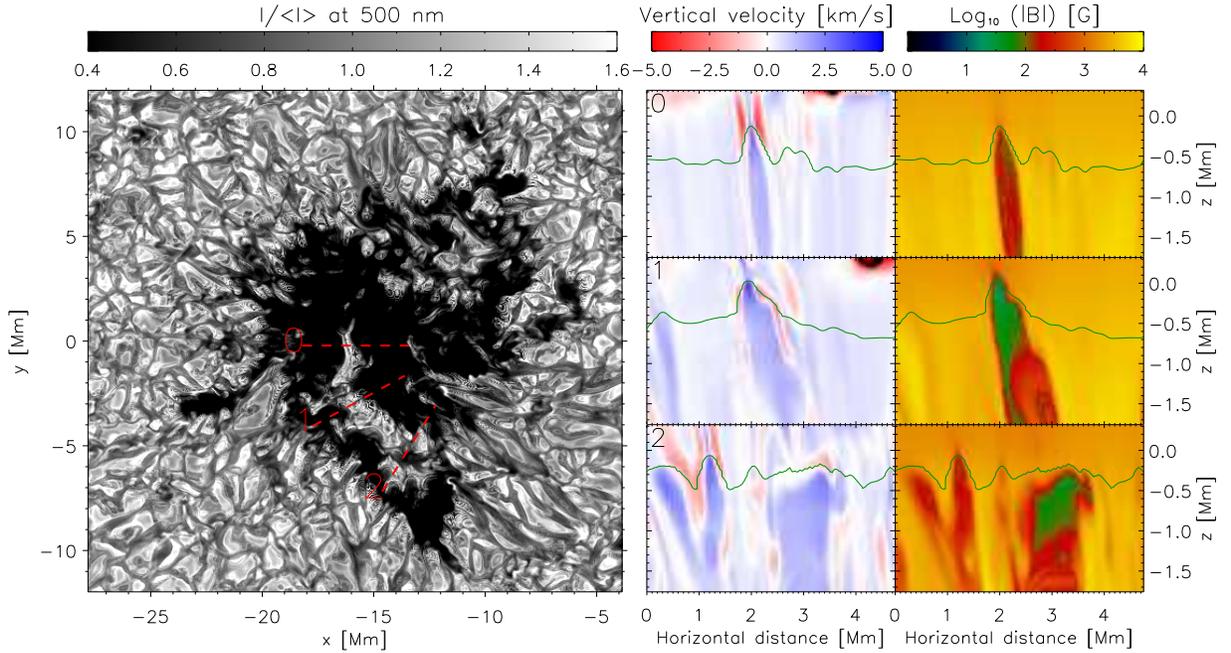}
\caption{Structure of a light bridge - The left panel shows the continuum brightness of one of the spots in the simulated active region ($t=15.4$ hrs). The formation of this light bridge is due to buoyant material entrained between adjacent regions of high magnetic field strength. Vertical cross-sections of the vertical velocity and magnetic field strength are shown for three cuts across segments of the light bridge of varying width. The green lines in the cross-sectional panels indicate the $\tau_{500} = 1$ surface.}\label{fig12}
\end{figure*}

As coalescence of flux concentrations progresses, ambient material with relatively weak field can become entrained in between adjacent high field-strength concentrations. An extended lane of entrained material trapped between neighboring regions of strong field appears as a light bridge in intensity maps. Figure~\ref{fig12} shows an example from the simulation. In this case, the horizontal convergence of strong field concentrations (manifested as dark umbrae in continuum intensity) traps ambient material with weak field to form a light bridge. Three vertical cross-sectional cuts oriented perpendicular to the light bridge show the vertical velocity distribution and magnetic field strength along parts of the bridge. At its end, located deep inside the spot ($[x,y]=[-18,0]$ Mm), the light bridge has a width of a few hundred km. The corresponding cross-sectional plots of $v_z$ and $|B|$ show that it is associated with an upflow of plasma with relatively weak magnetic field (a few hundred G at maximum). The upwelling locally lifts the $\tau_{500}=1$ surface by about $300$ km relative to the adjacent umbra. The expansion with height of the neighboring magnetic field pinches the upflow to force a cusp-like structure near $\tau_{500}=1$~\citep{Nordlund:LightBridge}. This is consistent with the observational study of light bridges by~\citet{Lites:3DStructureOfAR}. Just below $\tau_{500}=1$, the pinching accelerates the upflow from $0.5-1$ km s$^{-1}$ to a few km s$^{-1}$. Above $\tau_{500}=1$, the upflow speed decreases. This stagnation-like point partially traps low entropy material (radiatively cooled) and leads to a central dark lane within the light bridge.~{The slower speeds at the surface are consistent with the observational results by~\citet{Rimmele:MagnetoconvectionLightBridge}}. Cool material escape by feeding the downflows flanking the upflow. All these processes are similar to those occurring in simulations of umbral dots~\citep*{Schuessler:UmbralConvection}.

Vertical cross-sections across wider segments of the light bridge (Fig.~\ref{fig12}) tell a similar story. The vertical cross-section given in the middle panel with weak field, also flanked by a pair of periphery downflows. At an even wider section of the bridge with a width of a few Mm, the convective flow consists of a few overturning cells instead of a single one.

As the strong field concentrations continue to converge, the light bridge eventually disappears (at around $t=17$ hrs) by means of outflows channeled toward the edge of the spot. However, until the time when we halted the simulation ($t=37$ hrs), transient light bridges and umbral dots continued to appear in the pair of spots due to the tendency of entrained material to attempt to escape. Towards the later stages of the simulation, however, the frequency (and size) of light bridges diminished together with the amount of entrained material available for light bridge formation.

Together with previous models of umbral convection~\citep*{Schuessler:UmbralConvection} and sunspot simulations~\citep{Heinemann:PenumbraFineStructure,Rempel:PenumbralStructure,Rempel:SunspotStructure} suggest that umbral dots, penumbral filaments and light bridges are all manifestations of overturning magnetoconvection.

\section{Discussion}

In this paper, we presented the first radiative MHD simulation of the birth of an AR on the solar surface. To mimic the emergence of magnetic flux from the convection zone to the photosphere, a magnetic semitorus was kinematically advected upward through the bottom boundary, which is located at $7.5$ Mm below the photospheric base. Here is a list of interesting findings regarding the physics of how AR formation occurs in the simulation.

The magnetic field strength $B$ of the rising plasma in the emerging flux region roughly follow the scaling relation $B \propto \varrho^{1/2}$, where $\varrho$ is the mass density. This scaling relation is consistent with the scenario that rising plasma preferentially expand in the horizontal directions. This horizontal expansion disperses field over a large horizontal area. As time progresses, the dispersed magnetic fragments coalesce into gradually larger concentrations (in terms of flux content). When compact magnetic concentrations attain a flux of $\sim 10^{19}$ Mx or more, they appear as dark pores in intensity images. 

In order for dispersed field to become compact again, excess mass must be removed from the original emerging structure. This is facilitated by convective downflows, which act on horizontal field to form U-loops anchored below the surface. Magnetic reconnection within these loops transports mass away from rising field lines. This physical mechanism is the same as the one suggested by~\citet{Lites:SpaceScienceReviews} for explaining how the mass is discharged from the rising magnetic field in observed emerging flux regions.

~{Following the removal of mass from the original magnetic structure, magnetic flux must migrate into two concentrated trunks to form spots.} It is found that the azimuthally-averaged flows around the developing magnetic concentrations (corresponding to spots) contribute to the erosion of magnetic flux. However, correlations between the fluctuations of velocity and magnetic fields result in net inward migration of magnetic flux with the same sign as that of the spot.~{Such correlations result from the counter-streaming motion of opposite polarity fragments. This re-organization of the dispersed flux fragments is due to the presence of coherent subsurface magnetic roots, which influence the surface dynamics via the Lorentz force. In this sense, our result is akin to the tethered-balloon model of sunspot formation suggested by~\citet{Spruit:ClusterModel}.} We emphasize that these streaming magnetic polarities are more akin to the moving dipolar features~\citep[MDFs,][]{Bernasconi:EmergingFlux} found in emerging flux regions~\citep{StrousZwaan:HorizontalDynamics,Schlichenmaier:PenumbraFormation} than the so-called Type I moving magnetic features (MMFs), which likely also have serpentine field line geometries~\citep{SainzDalda:SeaSerpent,Kitiashvili:SeaSerpentPenumbra} but which may be associated with the decay of sunspots~\citep*{Kubo:MMFs}.

As the dispersed magnetic field collect together to form coherent spots, plasma with relatively weak field (at most a few hundred G) and high entropy may be entrained between regions of predominantly vertical field of a few kG strength. These strong field regions correspond to dark umbrae in intensity maps whereas the entrained weak field plasma in between manifest itself as a light bridge. Cross-sectional cuts of the vertical velocity and magnetic field strength distribution perpendicular to a light bridge in the simulation shares strong resemblance with similar cross-sectional cuts through umbral dots and penumbral filaments. This suggests a common convective origin for all three intensity features.

In this paper, we have focused our attention on analyzing how physical processes occurring in our radiative MHD simulation allow for the formation of an AR with certain observational properties (e.g. elongated granules and mixed polarity patterns in the emerging flux region, pore formation, transient light bridges etc). In a follow-on paper, we intend to carry out more detailed comparisons with observations of emerging flux regions. Such an exercise would be especially timely considering the rise of solar activity and the availability of vector magnetograms from both the Helioseismic Magnetic Imager onboard the Solar Dynamics Observatory and from the Hinode Solar Optical Telescope.

\acknowledgements
This work was made possible by NASA's High-End Computing Program. The simulation presented in this paper was carried out on the Pleiades cluster at the Ames Research Center. We thank the Advanced Supercomputing Division staff for their technical support.~{This work was supported by NASA contract NNM07AA01C at LMSAL. M. Rempel is partially supported through NASA grant NNH09AK02I to the National Center for Atmospheric Research. NCAR is sponsored by the National Science Foundation.}


\end{document}